\documentclass[doublecol]{epl2}
\usepackage{graphicx,amsmath,amssymb,mathrsfs}
\usepackage{epsfig}

\title{Symbolic dynamics techniques for complex systems: Application to share price dynamics}
\author{Dan Xu and Christian Beck}

\institute{Queen Mary University of London, School of Mathematical Sciences, Mile End Road, London E1 4NS, UK}

\pacs{05.45.-a}{Nonlinear dynamics and chaos}
\pacs{05.45.Tp}{Time series analysis}
\pacs{05.40.-a}{Fluctuation phenomena}

\abstract{
The symbolic dynamics technique is well-known for low-dimensional dynamical systems
and chaotic maps, and lies at the roots of the thermodynamic formalism of dynamical systems. Here we show that this technique can also be successfully applied to
time series generated by complex systems of much higher dimensionality.
    Our main example is the investigation of share price returns in a coarse-grained way. A nontrivial spectrum of
    R\'enyi entropies is found. We study how the spectrum depends on the time scale of returns,
    the sector of stocks considered, as well as the number of symbols used for the symbolic description. Overall our analysis confirms that in the symbol space transition probabilities
    of observed share price returns depend on the entire history of previous symbols, thus emphasizing the
    need for a modelling based on non-Markovian stochastic processes. Our method allows
    for quantitative comparisons of entirely different complex systems, for example the statistics of
    symbol sequences generated by share price returns using 4 symbols can be compared with that of
    genomic sequences.}

\begin{document}
\maketitle

\section{Introduction}

The symbolic dynamics technique is a powerful method to describe trajectories of a dynamical system
in a coarse-grained way \cite{schloegl,sym1,sym2,sym3,sym4,sym5,sym6,sym7,sym8,sym9}. It has been applied to many low-dimensional maps exhibiting
chaotic or critical behavior. Nontrivial correlations in the dynamical system manifest themselves
in a nontrivial spectrum of R\'enyi entropies \cite{renyi1,renyi2,renyi3,schloegl,sym3,sym4} and other observables associated with the set of allowed symbol sequences
and their corresponding probabilities. This technique, which was very popular in the 80s and 90s when a lot
of research on 1-dimensional maps was done \cite{sym1,sym2,sym3,sym4,sym9}, has been revived in more recent work
and applied in a more general context \cite{mod1,mod2,mod3,mod4}.

Our main aim in this paper is to illustrate that
the symbolic dynamics technique borrowed from dynamical systems theory can be successfully applied to
time series generated by general complex systems in higher dimensions,
far beyond the original 1-dimensional chaotic map approach. Our main example in the following will be
share price dynamics, which is of course ultimately produced by a complex market and
trader dynamics in a high-dimensional phase space \cite{rapisarda}. It is well-known that financial time series exhibit multifractal features \cite{fin1,fin2,fin3,fin4}, but here we present a somewhat different approach to this problem based on the symbolic dynamics
technique.
We will investigate the correlations and complex behavior associated with discrete symbol
sequences generated from observed share price returns on various time scales. We will quantify this
by the calculation of the corresponding R\'enyi entropies \cite{schloegl,renyi1,renyi2,renyi3} for symbol sequences based on historical data sets of
share price returns for various sectors, both on long (daily) time scales and on
short time scales (minutes). It turns out that
the stochastic process of symbol sequences observed for real share price data exhibit non-Markovian character.
To characterize differences between various companies (or communities in a complex system context), we will introduce a R\'enyi difference matrix which compares R\'enyi entropies of different subsystems.
The method developed allows for quantitative comparisons of different complex subsystems,
or even different scientific problems
due to the encoding in the symbolic
dynamics space. For example, it is possible to quantitatively compare
the statistical properties of genomic sequences \cite{genome1,genome2,genome3} with those
generated by coarse-grained share price movements, although both problems come from completely
different areas of science.

\section{Symbolic dynamics for share prices}

Let us apply the symbolic dynamics technique, well known from dynamical systems theory,
to a coarse grained description of a given time series. This time series is assumed to be generated by a suitable
observable of a dynamically evolving complex system. Rather than being very theoretical,
we choose a concrete example: Share price evolution, as created by complex market structures and trader speculations.
Often,
one is interested only in a very basic question related to a given complex system. For the example
of share prices, this question is quite straightforward: Basically one is interested in whether
the share price of a particular company will go up or down, and this problem of course
also depends on the time scale considered.
To generate symbol sequences associated with a question of this type
 we first need to choose a suitable partition of the phase space (a generating partition in the
 case of low-dimensional maps).

\begin{figure}[ht]
\begin{center}
\includegraphics[scale=0.5]{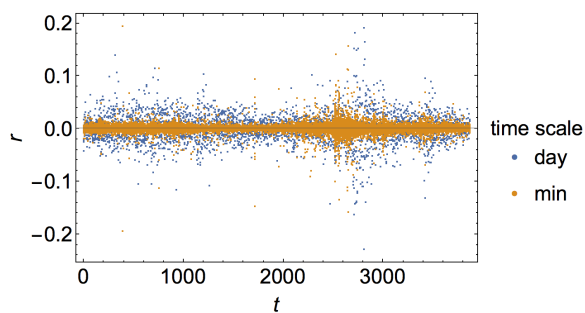}
\caption{Time series of log returns of a share price (in this example Alcoa Inc.) over the period 1998 to 2013, exhibiting intermittent outbursts
of strong volatility. The time unit is trading days. A year has about 251 trading days.}
\label{aaDr}
\end{center}
\end{figure}
 The easiest and most straightforward way for share prices is to consider just two symbols,
  i.e. to consider the possible values of logarithmic share price returns $\log \frac{S_{n+1}}{S_n}$
  in two disjoint subsets $A_1=(-\infty,0)$ and $A_2=(0,\infty)$, corresponding to negative and positive changes
  of the share price $S_n$ on a discrete time scale labelled as $n$. This is kind of a reduced phase space
  description for the question asked here.
  Note that $A_1$ and $A_2$ are chosen as open intervals. The point 0 (corresponding to no change)
    has measure 0, it does not influence the analysis. Of course, for other complex systems/time series
    one can define the subsets generating the symbol sequences differently, that depends on the problem and the question asked
    about the complex system. In general, much more complicated generating partitions arise in this way \cite{sym8,mod3}.
    %is not under consideration. This is because for very small time scale price data, say minutely recorded prices, due to the precision problem, a rounded-off price will somehow stay unchanged until the same value is caught again by the next point in time, hence, we eliminate the effect of zero changes on the probability measure for both subsets.

 Let's now
   look at an example data set of daily stock prices $S_n$ of the company
    Alcoa Inc. that covers the period January 1998 to May 2013 (Fig.~1).
    %For any given value (except 0) of a log return $\log \frac{S_{n+1}}{S_n}$ at a given time $n$, it has roughly equal chances to fall into either of the two sets $A_1$ and $A_2$.
    If the log return is an element of $A_1$, which is equivalent to  $S_{n+1}<{S_n}$, then we denote such a price decreasing event by the symbol $d$. Otherwise a price in $A_2$ stands for $S_{n+1}>S_n$ and is denoted by $u$ which means a price increasing. By this method we can attribute to the time series of share prices a symbol sequence
    $i_0, i_1, i_2,..., i_n,...$ where $i_n\in{\{d,u\}}$.
    We now consider subsequences of symbols of length $N$, where $N$ is small
as compared to the total number of data available. As we only have data for a limited set of data points, we divide up the whole
data sequence into $R$ segments of equal length $N$. Because each symbol has only two choices which are either $u$ or $d$, for any given $N$ we get up to
$ \omega (N)=2^N$
allowed subsequences.
Since the partition of the phase space is rather simple and our dataset is big enough to satisfy $R>>\omega (N)$, there will be many occurences that correspond to the same symbolic pattern. Hence we can easily acquire the probabilities of each allowed symbol sequences by determining the frequencies of how often the symbol sequence occurs
in the given data set. The probability of a given symbol sequence of length $N$ is then denoted as
$
p_j^{(N)} = p(i_0,...,i_{N-1})
$,
%
% & = \frac{N(i_0,...,i_{N-1})}{\sum_{i_0,...,i_{N-1}}{N(i_0,...,i_{N-1})}} \\
%  & = \frac{N(i_0,...,i_{N-1})}{R} \qquad j=1,..., \omega (N),
%\end{split}
%\end{equation}
where $j$ labels all possible sequences $i_0,..., i_{N-1}$.

\begin{figure}[!]
\begin{center}
\includegraphics[scale=0.8]{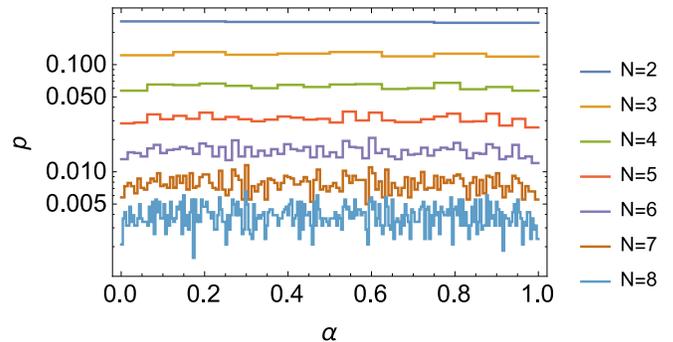}
\caption{Joint probabilities $p_j^{(N)}$ of the daily share prices movement dynamics for symbol sequence of length 2 to 8 for Alcoa Inc.}
\label{fig2}
\end{center}
\end{figure}

Now by encoding each allowed symbol sequence of length $N$ into a real number $\alpha$ on the unit interval using the
 binary expansion with $0=d$ and and $1=u$, we can produce a plot to visualise our probabilities. This means any given sequence of symbols $i_0, ..., i_{N-1}$ can be represented by a sequence of bits, in particularly we assign
\begin{equation}
x_{i_n}= \left\{
  \begin{array}{l l}
    1 & \quad \text{if }  i_n=u\\
    0 & \quad \text{if }  i_n=d
  \end{array} \right.
\end{equation}
where \(n=0,1,...,N-1\).
%We regard the encoded binary version $\textbf{x}^{(N)}=x_{i_0}, ...,x_{i_n}, ...,x_{i_{N-1}}$ of a $N$-step share price movement as a binary fraction(number after the decimal point), which can be eventually converted into a decimal fraction.
One can implement this by defining a coordinate $\alpha$ assigned to a given symbol sequence as
\begin{equation}
\alpha(\textbf{x}^{(N)})=\sum_{n=1}^{N}x_{i_{n-1}} 2^{-n}.
\label{phase01}
\end{equation}
Note that  $\alpha(\textbf{x}^{(N)}) \in{[0,1)}$.
In this way we allocate to each symbol sequence a real number $\alpha$ on the unit interval so that we can easily
visualize our results on the frequency of a given symbol sequence. This is shown in Fig.\ref{fig2}.

Note that numerically we restrict ourselves to $N\leq 8$ because of the fact that we only have a limited amount of data points. Larger $N$ would induce large stochastic
errors due to the fact that the statistics is not high enough to estimate the frequency of a given symbol sequence in a reliable way.

Also note that while for the simple example considered here
all symbol sequences are possible,  for more general
complex systems (and more complicated questions asked that are then encoded into the symbols,
 such as the problems in \cite{mod1,mod2,mod3,mod4}) there will
be a set of allowed and forbidden sequences in the symbol space. In this case the distribution in Fig.~2 will have gaps
and generally there will be a multifractal structure. Still everything that we define in this paper
can still be done in an analogous way.
%----------------------------------------------------------------------------------------------
\section{Information contents of the symbolic sequences}
   Given experimentally determined probability distributions with possible selfsimilar features such as in Fig.~\ref{fig2},
      it is meaningful to find a proper way to measure multifractal features.
For this purpose we use the well-known concept of \textit{R\'enyi information} \cite{schloegl,renyi1,renyi2,renyi3} defined as
\begin{equation}
I_{q}(p^{(N)})=\frac{1}{q-1} \ln \sum^{\omega(N)}_{j=1}(p_{j}^{(N)})^{q}.
\label{renin}
\end{equation}
Here $q$ is a real number and $\omega(N)$ is the number of allowed symbol sequences for a given $N$
(an allowed symbol sequence is one satisfying  $p_j^{(N)}\not= 0$).
 The index $j$ labels the various sequence probabilities.
The R\'enyi information measure can be regarded as a generalisation of the Shannon information, as for $q\to1$ we have
\begin{equation}
\lim_{q\to1}I_q(p^{(N)})=\sum_{j=1}^{\omega(N)}p_j^{(N)}\ln p_j^{(N)}=I(p^{(N)})
\label{shin}
\end{equation}
where $I(p^{(N)})$ denotes the Shannon information.

An important special case in Eq.\ref{renin} is the choice $q=0$,
\begin{equation}
I_0(p^{(N)})=-\ln \omega(N),
\end{equation}
which means $I_0(p^{(N)})$ decreases in a logarithmical way with the number of allowed symbol sequences.
Note that when we have a limited length $N$ of symbol sequences, Eq.\ref{renin} yields a finite value as $\omega (N)$ is finite.

Recall that for a given $N$ any symbol sequence is mapped onto a point of $[0,1)$,
 with equal distance between neighboring points.
    We call the distance of two neighboring coordinates at a given level
 in Fig.~{\ref{fig2} \textit{box size} and denote it by $\varepsilon$. In our case the box size is determined by $N$ as
$
\varepsilon=\frac{1}{2^N}$.
That means $\varepsilon$ is getting smaller as $N$ grows and when $N\to\infty$, $\varepsilon$ is approaching 0, in which case the R\'enyi information defined by Eq.\ref{renin} diverges. It is then useful to define the \textit{R\'enyi dimension} which is a useful quantity as it stays finite in the limit $\varepsilon \to 0$:
\begin{equation}
D(q)=\lim_{\varepsilon \to 0} \frac{I_q(p^{(N)})}{\ln \varepsilon}=\lim_{\varepsilon \to 0} \frac{1}{\ln \varepsilon} \frac{1}{q-1} \ln \sum^{\omega(N)}_{j=1}(p_{j}^{(N)})^{q}.
\end{equation}
An example of (finite box-size) R\'enyi dimensions evaluated for the data given in Fig.~2
    is shown in Fig.~\ref{fig3}.

\begin{figure}
\begin{center}
\includegraphics[scale=0.5]{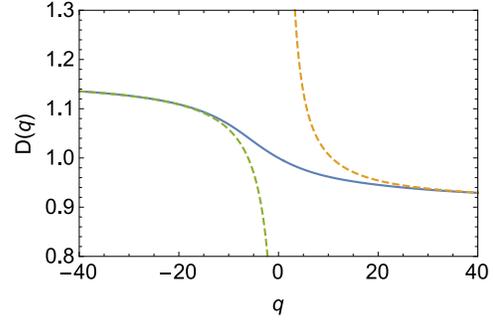}
\caption{R\'enyi dimensions together with the upper(orange dashed line) and lower(green dashed line) bounds for daily share prices movement of shares of Alcoa Inc (as numerically obtained for $N=8$).}
\label{fig3}
\end{center}
\end{figure}

     Of course the size of any data set is limited, in our case the smallest $\varepsilon$ that can be achieved
     with reliable non-fluctuating results is $\frac{1}{2^8}$.
     Given the finiteness of the data set it is useful to check some rigorous upper and lower bounds and
     monotonicity properties of the Renyi dimensions, valid for arbitrary probability measures.
     The R\'enyi dimensions are monotonically decreasing and their values must be positive for all $q$. Also, if all symbol sequences are allowed then we must get the value 1 when $q=0$. This is because
\begin{equation}
D(0)=\lim_{\varepsilon \to 0}\left( -\frac{1}{\ln \varepsilon}\right) \ln \sum^{\omega(N)}_{j=1}1=\lim_{\varepsilon \to 0} \left( - \frac{\ln 2^N}{\ln \frac{1}{2^N}}\right)=1.
\end{equation}

In addition, as shown in \cite{renyi2} there is a restriction of possible values of R\'enyi dimensions as a general upper and lower bound dimensions can be proved:
\begin{equation}
\frac{q'-1}{q'}D(q')\geq\frac{q-1}{q}D(q)   \qquad   \text{for } q'>q, q'q>0.
\label{ul}
\end{equation}
If we substitute $+\infty$ and $-\infty$ for $q$ in Eq.\ref{ul}, we obtain the upper bound
\begin{equation}
D(q)\leq\frac{q}{q-1}D(+\infty)  \qquad   \text{for } q>1
\label{upper}
\end{equation}
and a lower bound is given by
\begin{equation}
D(q)\geq\frac{q}{q-1}D(-\infty)  \qquad   \text{for } q<0.
\label{lower}
\end{equation}

   We have checked these bounds for our data set with $N=8$, the result is also shown
    in Fig.~\ref{fig3}. Clearly the inequalities (\ref{upper}) and (\ref{lower})
     are satisfied by our experimental data.

From the R\'enyi dimensions $D(q)$ one can proceed to the singularity spectrum $f(\tilde{\alpha})$
of crowding indices $\tilde{\alpha}$
by a Legendre transformation in the thermodynamic
formalism. However, as the information contained
in $f(\tilde{\alpha})$ spectra is the same as the one in the generalized dimensions $D(q)$, we will not further proceed along these lines here,
but refer the reader to suitable literature introducing to this topic \cite{halsey,schloegl}.
%-----------------------------------------------------------------------
\section{R\'enyi entropies}

\begin{figure}[ht]
\begin{center}
\includegraphics[scale=0.8]{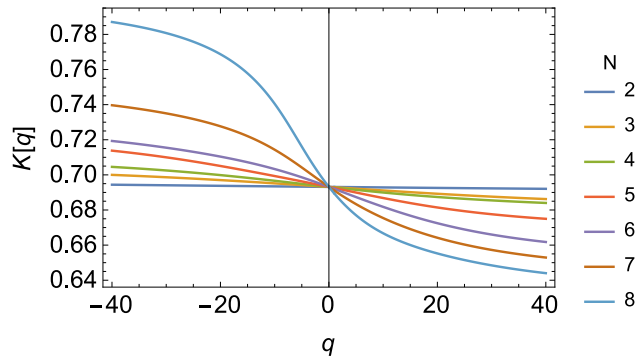}
\caption{R\'enyi entropies for the daily share prices movement dynamics for symbol sequence of length 2 to 8 for Alcoa Inc.}
\label{fig4}
\end{center}
\end{figure}

Unlike the R\'enyi dimension, which generally is a property
of a given multifractal probability measure and which does not {\em a priori} contain
any dynamical information, the \textit{R\'enyi entropy} $K(q)$ measures the production (or loss) of information
encoded in the symbol sequences. This is the quantity of direct interest for share price evolution.
The R\'enyi entropy $K(q)$ is defined in the limit $N \to \infty$ and given by
\begin{equation}
K(q)=\lim_{N\to\infty}\frac{-I_q^{(N)}}{N}=\lim_{N\to\infty}\frac{1}{1-q}\frac{1}{N}\ln\sum^{\omega(N)}_{j=1}(p_j^{(N)})^q
\end{equation}
Of course, by projecting symbol sequences onto points on the real line (as done in Fig.~\ref{fig2}), both
approaches can be made formally equivalent, but the dynamical information is then just encoded in
a suitable multifractal measure.

   Fig.~\ref{fig4} shows finite-$N$ versions of R\'enyi entropies for share price returns where $N$ grows from 2 to 8.
    The Renyi entropies have of course the same dependence on $q$ as the Renyi dimensions of
    the corresponding multifractal measure that encodes the symbol sequence probabilities on the unit interval.
%\begin{figure}[ht]
%\begin{center}
%\includegraphics[scale=0.8]{sym-fig4.eps}
%\caption{R\'enyi entropies for the daily share prices movement dynamics for $N=2,...,6$ for Alcoa Inc.}
%\label{s2aaRenEnDr}
%\end{center}
%\end{figure}

For $q=0$ we get
\begin{equation}
K(0)=\lim_{N\to\infty}\frac{1}{N}\ln2^N=\ln2,
\end{equation}
which is the topological entropy.
Moreover for $q\to1$, using Eq.~\ref{shin},
\begin{equation}
K(1)=\lim_{N\to\infty}\frac{-I(p^{(N)})}{N}
\end{equation}
    we get the \textit{Kolmogorov-Sinai entropy} asssociated with the symbol sequences of share price changes.
    Again one can proceed from the function $K(q)$ to an equivalent spectrum of dynamical indices $g(\tilde{\alpha})$ by
    Legendre transformation (see, e.g. \cite{schloegl}).
%-----------------------------------------------------------------------------
\section{Small time scales}
We have quantified the joint probabilities of the daily share price movements by the corresponding R\'enyi entropies. We are now interested in how these observables depend on the time scale of the symbolic dynamics.
Instead of using the daily share prices, the now analysed data set consists of share prices
recorded each minute for the same period from 1998 to 2013; this covers about 1.5 million data points. We repeat the same analysis as before and consider symbol sequences up to length 8. By using the same partition method as in the previous section we obtain the multifractal probability distributions of symbol sequences as shown in Fig.~\ref{s2aaProMr}.

\begin{figure}[!]
\begin{center}
\includegraphics[scale=0.8]{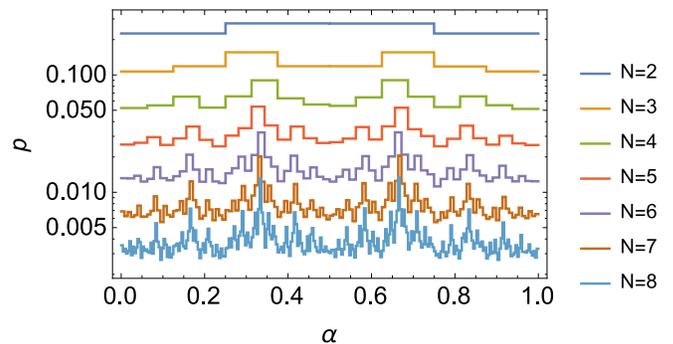}
\caption{Joint probabilities of share price dynamics for symbol sequence of length 2 to 8 for Alcoa Inc.,
evaluated on a time scale of minutes.}
\label{s2aaProMr}
\end{center}
\end{figure}

Compared with Fig.~2, the probability distributions on a small time scale are significantly different from those on the daily time scale. Note that there are some local maxima which are reproduced in a self-similar way.
While these densities are non-smooth, the advantage of proceeding to the Renyi dimensions (or entropies) is that
in this way a smooth dependence on the scanning paramter $q$ is produced. This is shown in Fig.~6 and 7.

\begin{figure}[ht]
\begin{center}
\includegraphics[scale=0.8]{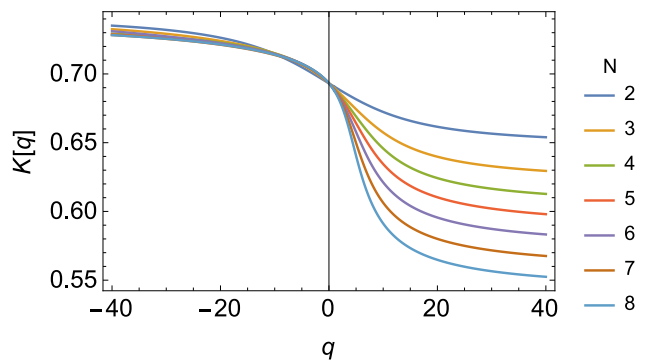}
\caption{The R\'enyi entropies as obtained for
Alcoa shares on a time scale of minutes, with $N=2,..., 8.$ }
\label{minute}
\end{center}
\end{figure}

We see, similar to the daily time scale, that both the R\'enyi entropies and the R\'enyi dimensions
 are monotonically decreasing with respect to the paramter $q$, with larger $N$
 generating a more pronounced $q$-dependence for positive $q$, whereas there is
 hardly any $N$-dependence for $q<0$.
                     Fig.~7 also shows the upper and lower bound.

\begin{figure}[ht]
\begin{center}
\includegraphics[scale=0.5]{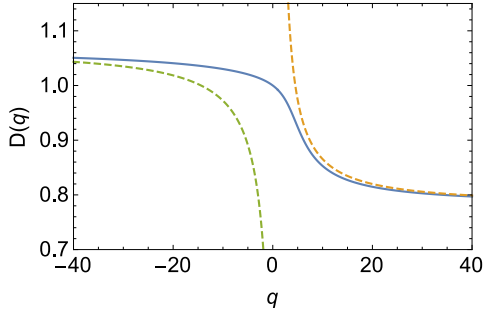}
\caption{Upper and lower bounds (dashed lines) for the R\'enyi dimensions (solid line) of
Alcoa shares on a time scale of minutes ($N=8$).}
\label{minute}
\end{center}
\end{figure}

An important property that we have checked for our data sets is the fact
that quite generally the conditional probabilities
\begin{equation}
p(i_N|i_0, \ldots , i_{N-1})=p(i_0, \ldots i_N)/p(i_0, \ldots i_{N-1})
\end{equation}
depend on the entire history $i_0, \ldots , i_{N-1}$, i.e. a Markovian model cannot capture
the complex features in the symbol space. For example, given a long alternating
sequence $u,d,u,d$ it is statistically slightly more likely to observe the next symbol as $u$.
%------------------------------------------------------------------------------------------------------------
\section{4-symbol partitions}
    So far we used the simplest method to study the symbolic dynamics of the share price movements,
    just considering whether a share price goes up or down. We may, however, also ask
    a more detailed question, such as whether the share price goes up slightly or strongly.
     To detect further details, we may generate a refined version of the phase space partition $\mathbf{A}=\{A_1, A_2\}$ where previously $A_1=(-\infty, 0)$ and $A_2=(0, \infty)$. Assume there exists a real number $c$ where the log returns have equal 1-point probabilities to fall into each element of a partition $\mathbf{B}=\{B_1, B_2, B_3, B_4\}$ given by
\begin{equation}
B_1=(-\infty, -c), B_2=(-c, 0), B_3=(0, c), B_4=(c, +\infty).
\end{equation}
In other words, this partition is chosen in such a way
that the 1-point probabilities $p(B_i), i=1,2,3,4$ of the log returns lying in each set $B_i$ are identical to $1/4$. For Alcoa Inc. share prices on daily and minutely time scales, we find $c$ is equal to 0.014 and 0.00088, respectively.
Instead of denoting the time evolution by $u$ and $d$, we re-define our symbol sequence by
\begin{equation}
i_0, i_1, ..., i_n, ...\qquad i_n\in\{b_1, b_2, b_3, b_4\}
\end{equation}
where $b_i$ corresponds to a log return in $B_i$. For a given length $N$, the number of
allowed sequences is $\omega(N)=4^N$.
We may also upgrade the previous approach to a 4-level symbolic sequence given by

\begin{equation}
x_{i_n}= \left\{
  \begin{array}{l l l l}
    0 & \quad \text{if }  i_n=b_1\\
    1 & \quad \text{if }  i_n=b_2\\
    2 & \quad \text{if }  i_n=b_3\\
    3 & \quad \text{if }  i_n=b_4,
  \end{array} \right.
\end{equation}
A symbol sequence of length $N$ can be encoded as a coordinate $\alpha$ on the unit interval based on the representation
\begin{equation}
\alpha(\textbf{x}^{(N)})=\sum_{n=1}^{N}x_{i_{n-1}} 4^{-n},  \quad \alpha(\textbf{x}^{(N)})\in{[0,1)}.
\label{phase0123}
\end{equation}
  The plot of joint probabilities in the case of an alphabet of 4 symbols is shown in Fig.~8,
  the corresponding finite-$N$ R\'enyi entropies are shown in Fig.~9.
\begin{figure}[ht]
\begin{center}
\includegraphics[scale=0.8]{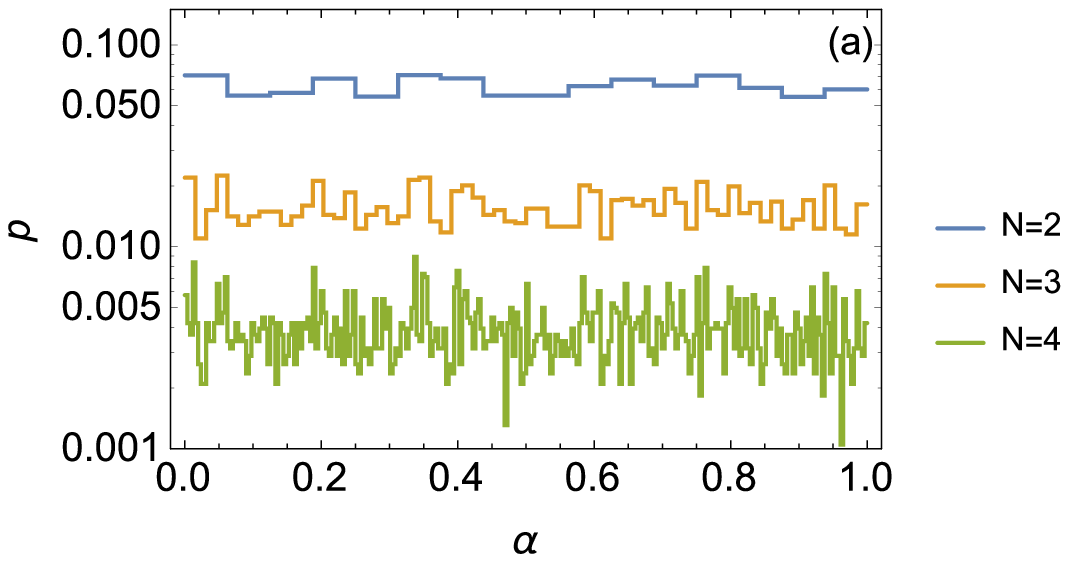}
\includegraphics[scale=0.8]{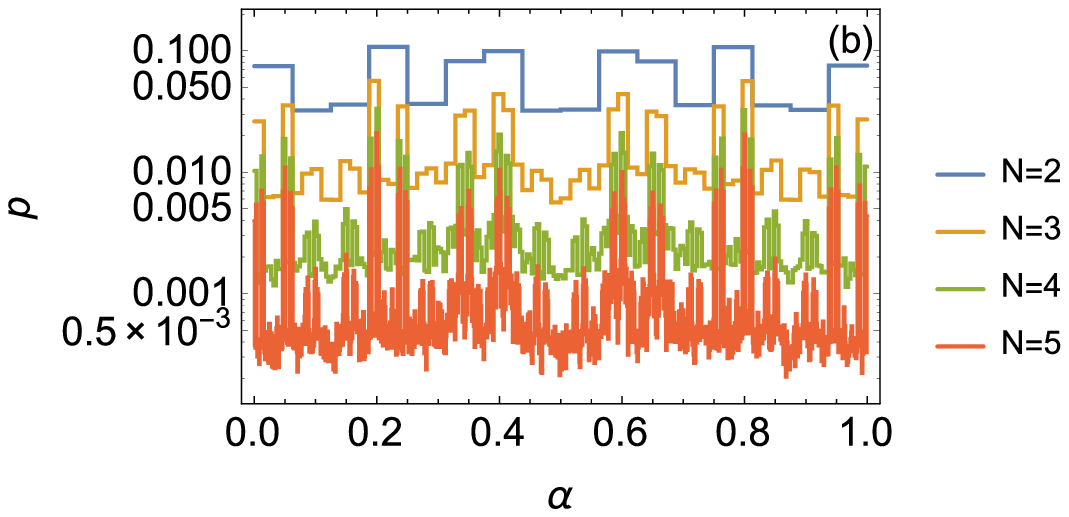}
\caption{Symbol sequence probabilities of length $N$ for Alcoa shares using an alphabet of 4 symbols (daily (a) and minute (b) time scale).}
\label{s4aaProDr}
\end{center}
\end{figure}

\begin{figure}[ht]
\begin{center}
\includegraphics[scale=0.7]{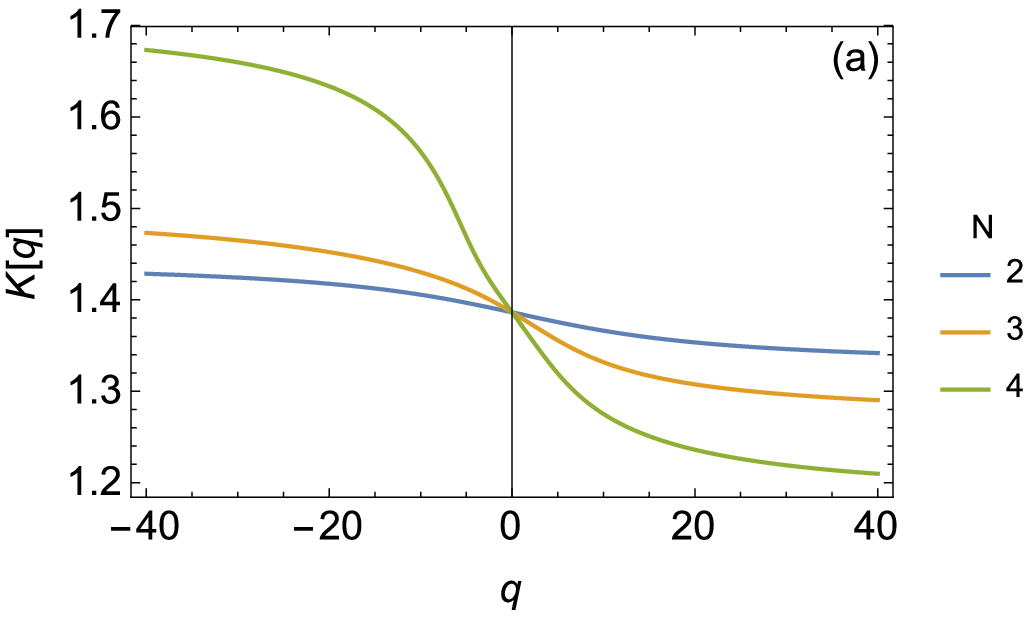}
\includegraphics[scale=0.7]{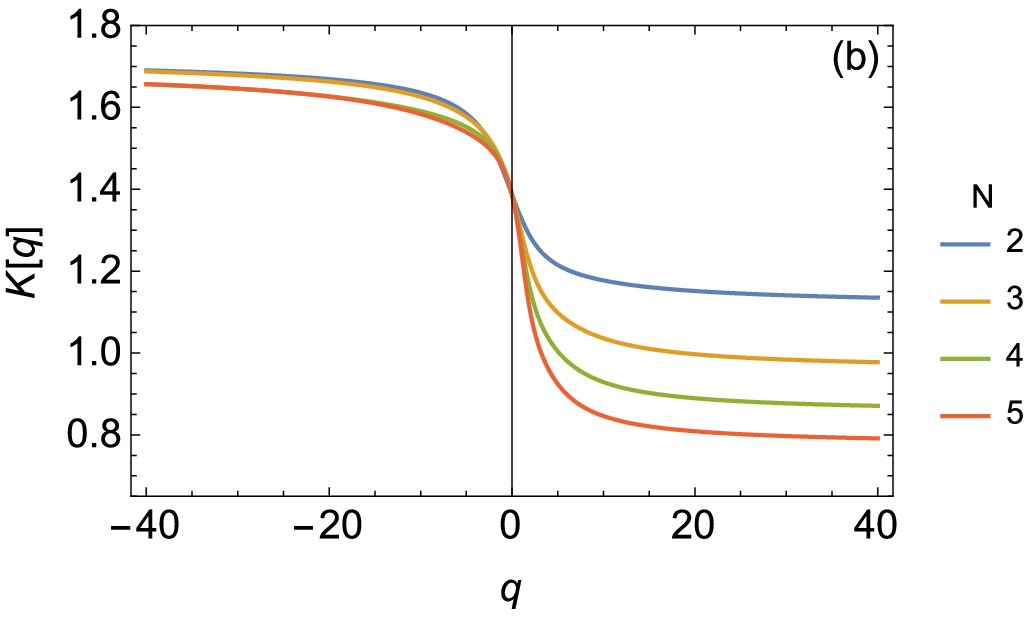}
\caption{Finite-$N$ R\'enyi entropies for the same data as in Fig.~8a,b.}
\label{s4aaRenEn}
\end{center}
\end{figure}

%------------------------------------------------------------------------------
\section{Different companies}
 So far we only looked at a particular example of a stock, Alcoa shares.
  The really interesting work on complex system analysis starts when one compares R\'enyi entropies of different
  companies, or even R\'enyi entropies of entirely different complex
  systems described by the same alphabet of symbols. Ultimately we really want to learn something
  about the complex market structure and dynamics represented by different companies, sectors, or communities
  for general complex systems. Still one can let each community generate a suitable time series for a suitable
  observable and then
  analyse this time series with the symbolic dynamics technique described so far. We are then interested
  in differences or similarities of the obtained Renyi entropies.

\begin{figure}[ht]
\begin{center}
\includegraphics[scale=0.7]{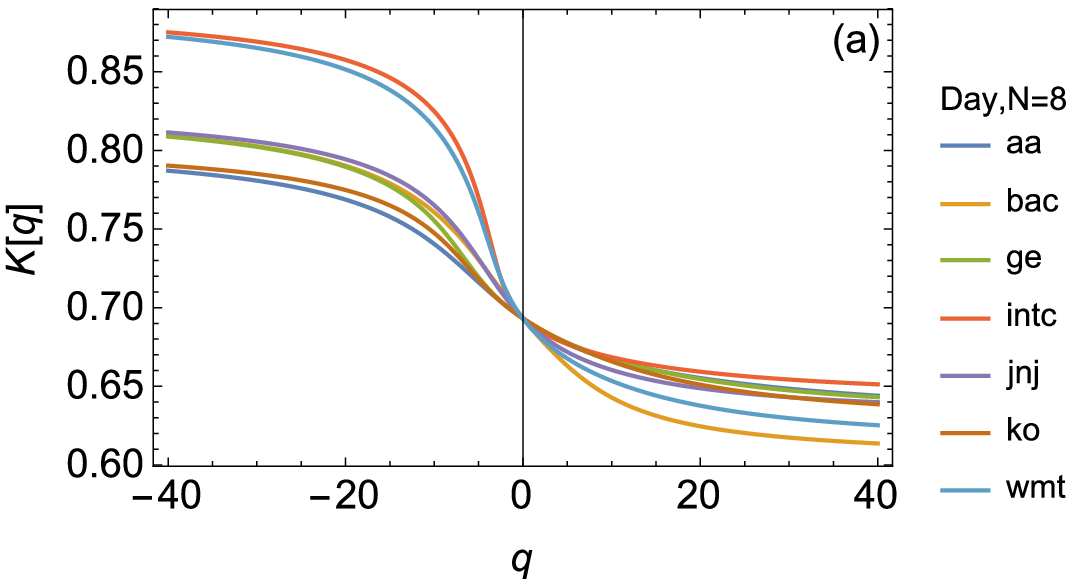}
\includegraphics[scale=0.7]{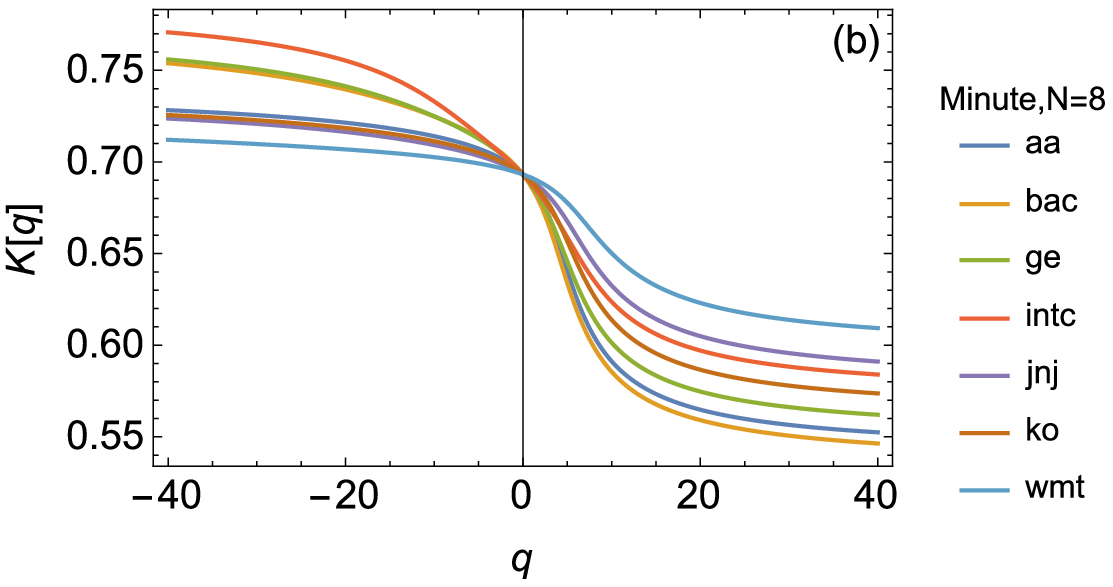}
\caption{R\'enyi entropies for 7 different companies on daily scale (a) and minute scale (b).
The alphabet contains 2 symbols.}
\label{7cd}
\end{center}
\end{figure}

      To illustrate this we looked at symbol sequences as generated by share prices of 7 different companies,
      Alcoa (aa), Bank of America (bac), General Electric (ge), Intel (intc), Johnson \& Johnson (jnj), Coca Cola (ko),
      WalMart (wmt), representing the sectors basic materials, financial, industrial goods, technology, healthcare, consumer goods, services.
               The results are shown in Fig.\ref{7cd}.
               It can be observed that bac
                 appears to have the lowest R\'enyi entropy in the region $q>0$ as compared to other stocks,
                 overall the $q$-dependence for financial stocks is most pronounced. This could have to do
                    with the fact that financial stocks have relatively strong fluctuations and exhibit nontrivial correlations,
                    described by a non-trivial spectrum of R\'enyi entropies. In any case, different companies are
                    characterized by a different spectrum of R\'enyi entropies both on a daily (Fig.~10a) and
                    minute (Fig.~10b) time scale. The smooth dependence on the parameter $q$ can be used for an
                    effective thermodynamic description of the complex behavior involved,
                    with different emphasis given to low and high proababilities depending on the
                    value of the scanning parameter $q$.

\section{Quantifying similarities in the symbol sequence statistics}

We may now wish to compare in a quantitative way how much the R\'enyi entropies of
different companies (or communities in the general complex system context) differ.
For this purpose we define a {\em R\'enyi difference matrix} $ R_{ij}$ as follows:
\begin{equation}
R_{ij} = \frac{1}{q_{max} -q_{min}} \int_{q_{min}}^{q_{max}} |K_i(q)-K_j(q)|^\kappa dq
\end{equation}
Clearly, if two companies $i$ and $j$ have the same statistics of symbol sequences,
described by the same functional dependence $K_i(q)=K_j(q)$, then the R\'enyi difference matrix element is
$R_{ij}=0$. Otherwise, the entry $R_{ij}$ integrates up differences in the R\'enyi entropy spectra,
and averages them over $q$, weighted with the parameter $\kappa$.

\begin{figure}[ht]
\begin{center}
\includegraphics[scale=0.15]{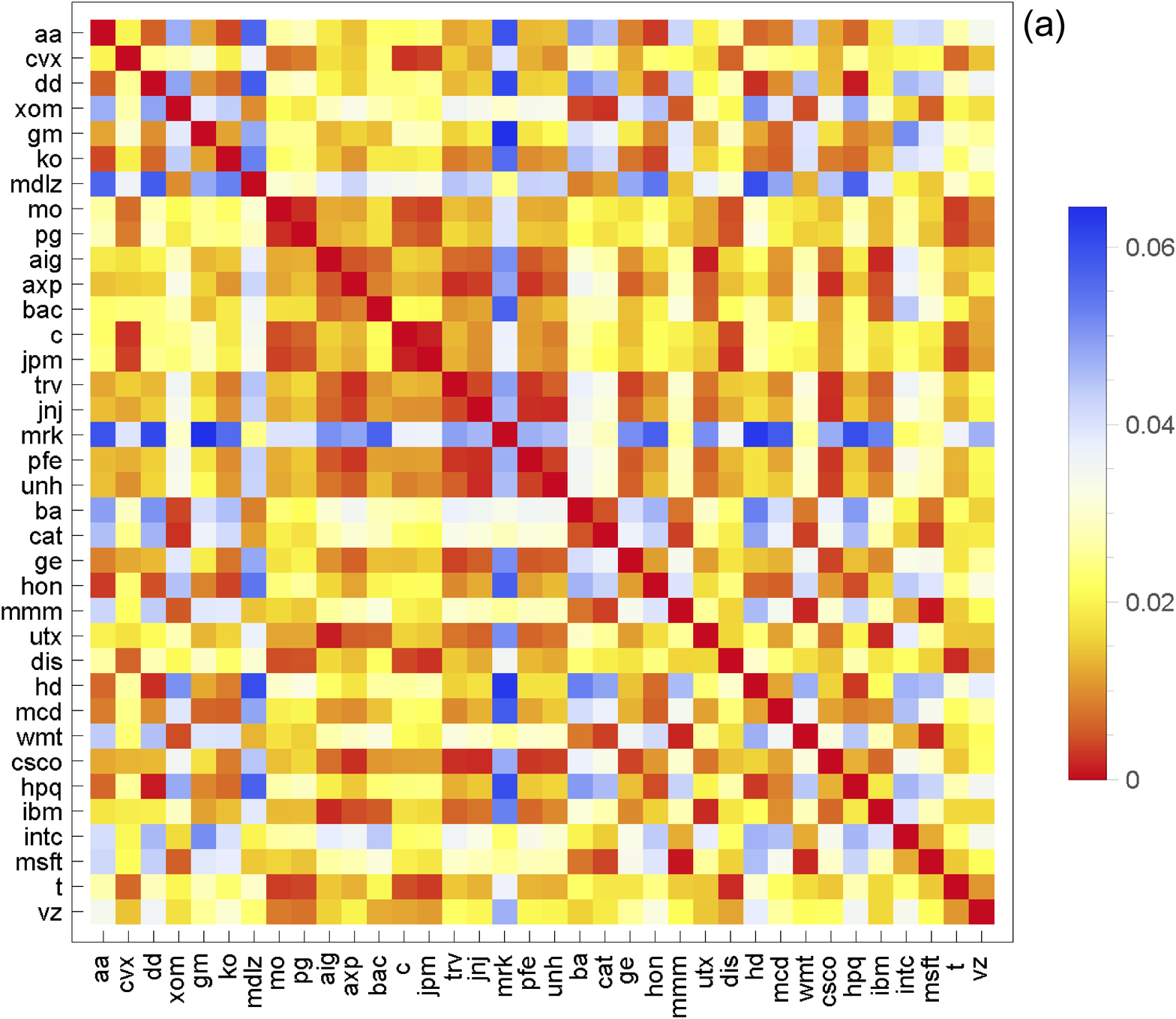}
\includegraphics[scale=0.15]{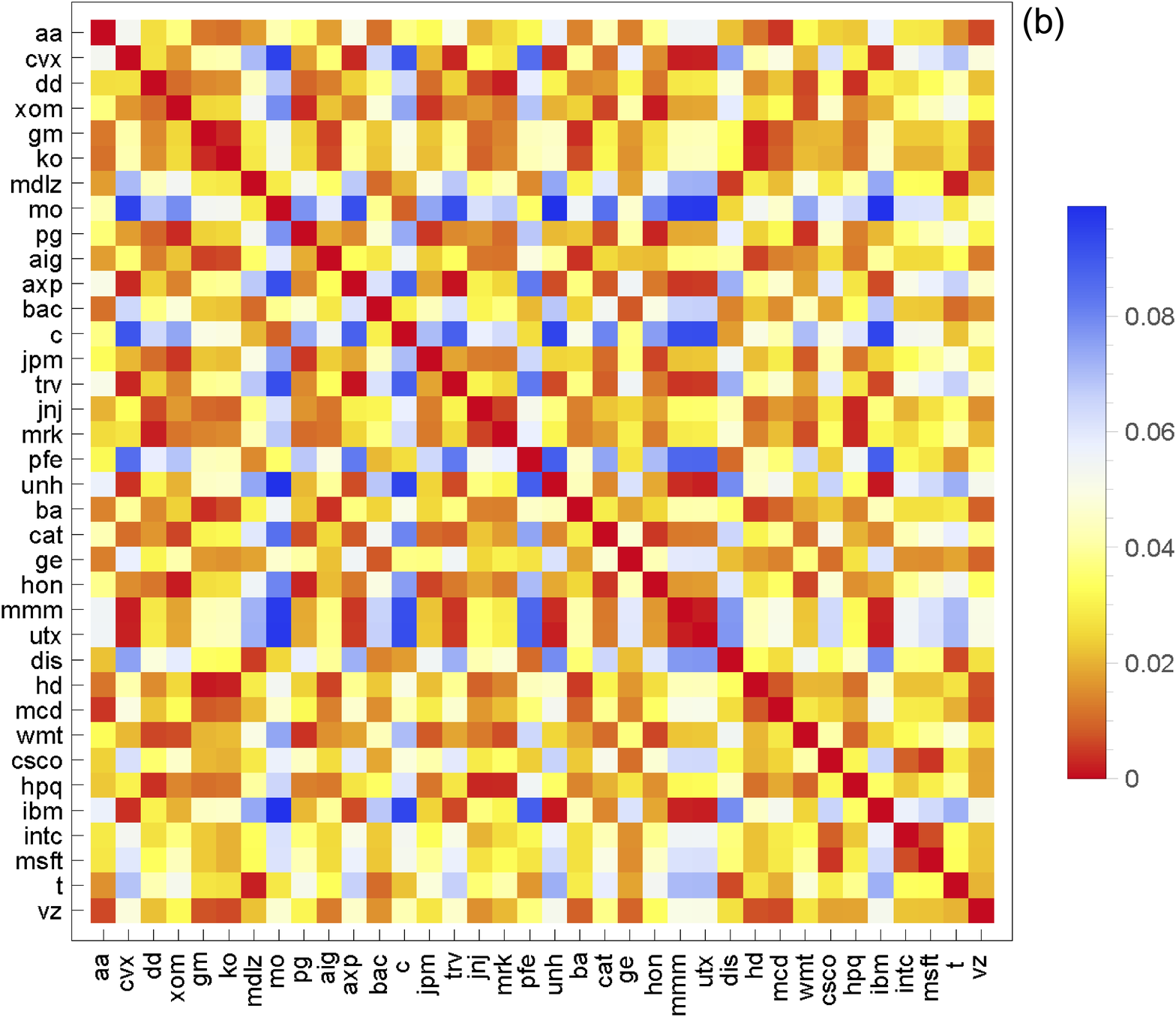}
\caption{R\'enyi difference matrix $R_{ij}$ on a daily time scale (a) and minute time scale (b) for 36 stocks
traded at the NYSE-Nasdaq.
 The value of the parameter $\kappa$ was chosen as $\kappa =1$. Similar pictures arise for other values of $\kappa$.}
\end{center}
\end{figure}

Fig 11 shows a colour encoding of such a R\'enyi difference matrix.
For 36 different companies we evaluated $R_{ij}$, choosing
$\kappa =1$ and $q_{min}=-40, q_{max}=+40$.
The R\'enyi difference matrix allows one to
single out
major differences and similarities in the symbol sequence statistics of different companies/communities in a quantitative way. In our case, for example, the healthcare company Merck (mrk) is identified as
having an unusual
R\'enyi entropy spectrum on a daily scale, different from that of most other companies, visible here as a pronounced vertical and horizontal blue line in
the pattern generated. On the other hand, on the small time scale of minutes this company
is much more similar to the others.

An interesting final remark is at order. Once a suitable question has been asked about a complex system,
and a symbolic dynamics constructed, we can then compare different complex systems,
or subsystems thereof, whatever their origin,
as all relevant information is encoded in the symbol sequence statistics. To give an example, in \cite{genome2,genome3}
the complexity of DNA sequences of the human genome was investigated, by calculating the Renyi entropies
associated with the sequence statistics of the 4 bases  A,T,G,C. Once this function is obtained, it can then be compared
using the above R\'enyi difference matrix with other genomes, in just the same way as we compared
the differences between different companies in Fig.~11. But more drastically, we can even compare in a quantitative
way (via $R_{ij}$) the R\'enyi spectra of completely different complex systems, such as the complexity
of financial markets and the complexity of genomes. This is the advantage of the symbolic
dynamics encoding technique: Once a function $K_i(q)$ has been obtained, one can compare it
in a quantitative way with another functions $K_j(q)$,
whatever its origin, and thus measure differences in complexity and information production in a quantitative way.
%\begin{table}
%\tiny
%\centering
%\caption{Companies investigated in Figs 10}
%\label{table}
%\begin{tabular}{|l|c|c|}
%\hline
%\textbf{Company}                  & \textbf{Sector}  & \textbf{Industry}          \\ %\hline
%Alcoa Inc. (AA)                   & basic materials  & Aluminum                   \\ %\hline
%Bank of America Corporation (BAC) & financial        & Money Center Banks         \\ %\hline
%General Electric Company (GE)     & industrial goods & Diversified Machinery      \\ %\hline
%Intel Corporation (INTC)          & technology       & Semiconductor - Broad Line \\ %\hline
%Johnson \& Johnson (JNJ)          & healthcare       & Drug Manufacturers - Major \\ %\hline
%The Coca-Cola Company (KO)        & consumer goods   & Beverages - Soft Drinks    \\ %\hline
%Wal-Mart Stores Inc. (WMT)        & services         & Discount, Variety Stores   \\ %\hline
%\end{tabular}
%\end{table}

%----------------------------------------------------------------------------------
\section{Conclusions and outlook}

Although the examples considered in this paper were all based on symbol sequences generated by share price returns,
it is clear that the same method can be applied to symbol sequences generated by time series
of all kinds of complex systems, whatever their origin. In this way the R\'enyi entropies associated with such a symbolic description
allow for a quantitative comparison of the dynamical properties in symbol space, making it easy to compare
different complex systems, or different sub-structures/communities within
a given big complex system. In fact, one can in this way compare entirely different complex systems, for example the
R\'enyi entropies associated with share price changes (using an alphabet of 4 symbols) can be compared with those of
genomic sequences \cite{genome2,genome3} or those of successive quantum mechanical
measurements \cite{graudenz}. The most important dynamical information is then
encoded in form of the shape of the function $K(q)$, allowing the application of thermodynamic tools \cite{schloegl}.
In this way a quantitative comparison of different systems is possible, solely based on the R\'enyi information
contents of the coarse-grained symbolic description. The extension of the methods presented here to more
complicated symbolic dynamics generated by other types of complex systems is straightforward.

%----------------------------------------------------------------------------
%	REFERENCE LIST
%----------------------------------------------------------------------------------------
\bibliographystyle{plain}

\end{document}